%% file: ACM-publications.tex
\documentclass[times]{acm-book}

\usepackage{amsmath}    
\usepackage{graphicx} 
\usepackage[utf8]{inputenc}
\usepackage[ngerman,english]{babel}
\usepackage{stfloats}
\usepackage{pdfpages}
\usepackage{hyperref}

\begin{document}

\title{Handbook on Socially Interactive Agents}

%
%
%

%
\thispagestyle{empty}
\setboolean{@twoside}{false}
\setlength{\hoffset}{0cm}
\setcounter{page}{0}\includepdf[pages={1},pagecommand={},width=\paperwidth]{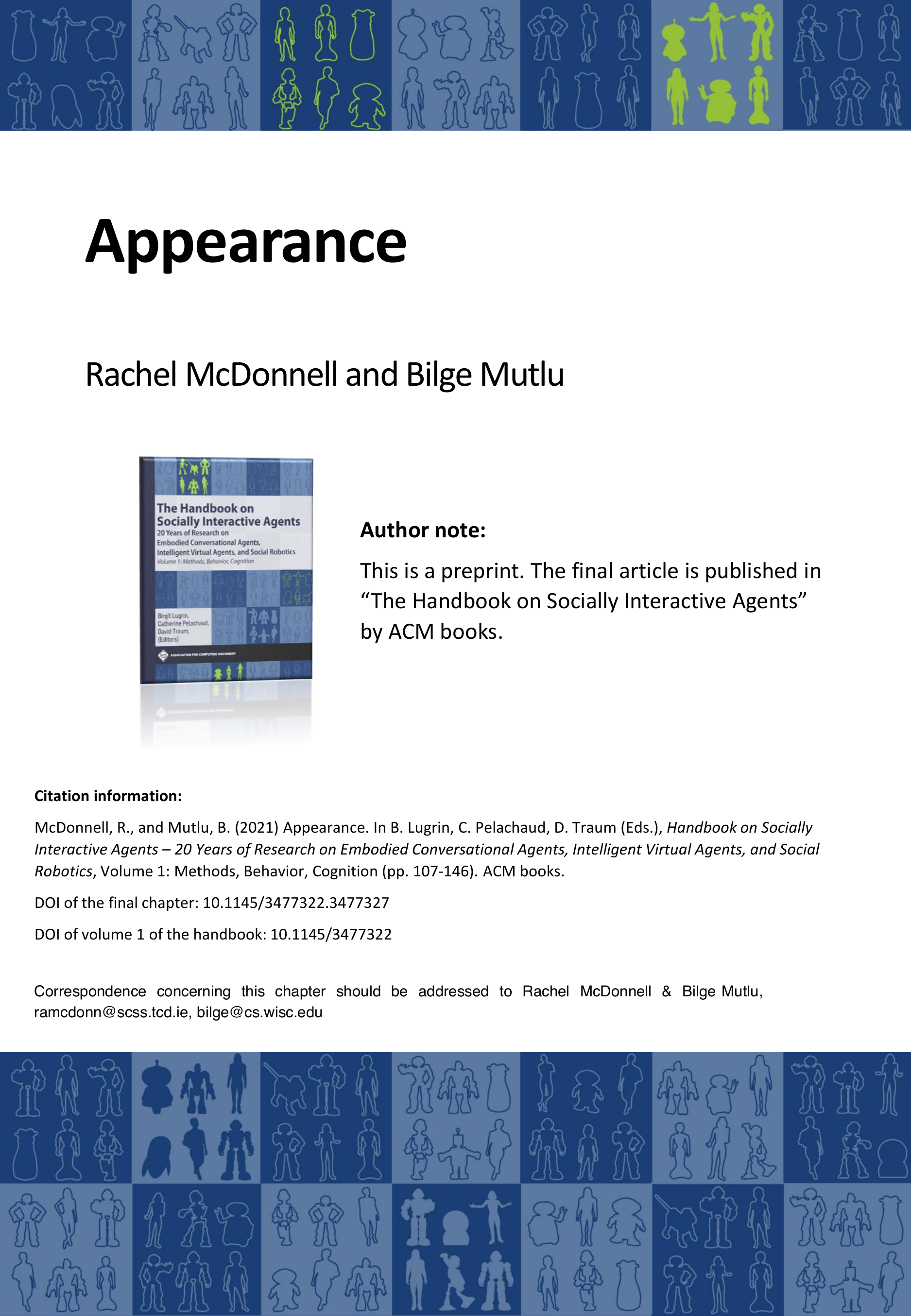}
\thispagestyle{plain}
\setlength{\hoffset}{.5in}
\include{II_appearance}
%
%
%
%
%

\backmatter
\bibliography{ACM-publications}

\end{document}

%% file: II_appearance.tex
\chapter{Appearance}
\chapterauthor{Rachel McDonnell, Bilge Mutlu}
 
\section{Why appearance?}
One might question why we need appearance for the metaphor to work, as voice assistants can effectively express characteristics of a metaphor solely through behavior. We argue that, although disembodied agents can effectively serve as computer-based assistants in specific scenarios of use, for example, involving driving and visually impaired users, appearance provides a ``locus of attention'' \cite{cassell2001embodied} for the cognitive and interactive faculties of the user of the system. Additionally, human communication mechanisms, such as mutual gaze, turn-taking, body orientation, necessitate the presence of appropriate visual cues to properly function, making appearance a necessity for agent design. Studies of human-human, human-agent, and human-robot interaction provide strong evidence that such mechanisms work more effectively when parties provide appearance-based cues. The mere presence of a form of embodiment in interacting with an agent improves social outcomes, such as motivation \cite{mumm2011designing}. As the scale and modality of appearance get closer to that of the metaphor, these outcomes further improve; human-scale and physical agents have more perceived presence \cite{kiesler2008anthropomorphic} and persuasive ability \cite{bainbridge2011benefits} than scaled-down and virtual agents.

\section{History}

\begin{figure}[t]
  \centering
  \includegraphics[width=\columnwidth]{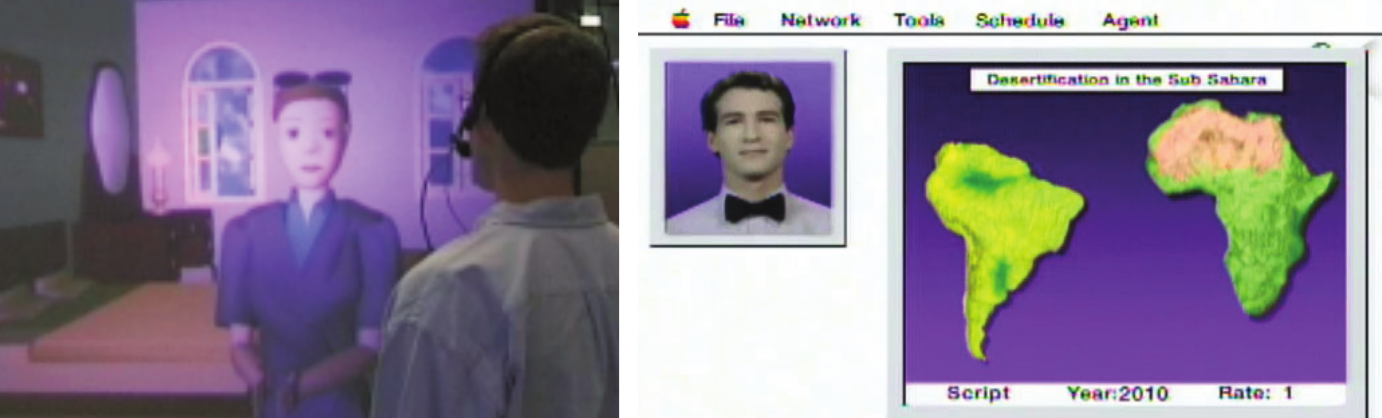}
  \caption{Early examples of virtual embodiments. \textit{Left:} The Rea real-estate agent \cite{cassell2000embodied}; \textit{Right:} the personal assistant envisioned for Knowledge Navigator \cite{Sculley:1989}.}
  \label{fig:example-virtual}
\end{figure}

Agents with virtual and physical embodiments follow different historical trajectories. Virtual agents, also called embodied conversational agents, a term coined by \citet{cassell2000embodied}, are ``computer-generated cartoonlike characters that demonstrate many of the same properties as humans in face-to-face conversation, including the ability to produce and respond to verbal and nonverbal communication.'' Early visions for virtual agents involved characters involved re-played recordings of human performers, such as the intelligent personal agent included in the Knowledge Navigator concept developed by Apple in 1987 \cite{colligan2011how} (Figure \ref{fig:example-virtual}). First implementations of virtual agents were stylized nonhuman or human characters that were generated through 3D modeling and rendering and were embedded within virtual environments. An example of early nonhuman characters included Herman the Bug, an animated pedagogical agent embedded within a virtual learning environment \cite{lester1997mixed}. Another early example is Rea, a real-estate agent that followed a stylized humanlike design and appeared within a simulated home environment \cite{cassell2000embodied}. Although these examples represent agents that are controlled and visualized by computer systems, the design of such nonhuman and human characters have a long history in shadow puppetry, dating back to the first millennium BC \cite{orr1974puppet}. These characters were designed for storytelling and entertainment, and the character designs reflected historical or cultural figures as well as characters developed with backstories. The design of the characters also include stylizations and ornamentations that reflect their ethnic and cultural context, such as the character Karagöz that followed a stylized human design with clothing and storyline from the 16-19th century Ottoman Empire \cite{scarce1983karagoz}.

\begin{figure}[!b]
  \centering
  \includegraphics[width=\columnwidth]{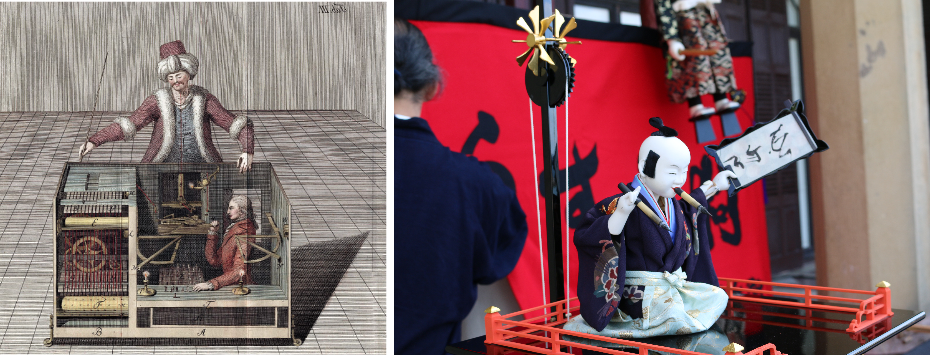}
 \caption{Early physical agents. \textit{Left:} Mechanical Turk automata by Joseph 
Racknitz (1789), image courtesy of Humboldt University Library; \textit{Right:} a tea-serving Karakuri puppet, Karakuri ningyo (c) 2016 Donostia/San Sebastian.}
  \label{fig:example-automata}
\end{figure}


The design of agents with robotic embodiments date back to mechanical humanoid automata designed as early as the 10th century BC \cite{hamet2017artificial}. As did with virtual characters and shadow puppetry, the physical appearance of these early automata also followed stylized humanlike forms. Examples, shown in Figure \ref{fig:example-automata}, include the design of the Mechanical Turk, a covertly human-controlled chess-playing machine that integrated a humanoid chess player on a wooden chest where the human operator hid \cite{simon1999enlightened}. Karakuri puppets, mechanical automata designed in the 17-19th century Japan to be used, for example, to ceremonially serve tea, followed a stylized humanlike appearance and traditional Japanese clothing \cite{yokota2009historical}. Although the appearance of robotic agents has overwhelmingly followed a human form with some level of stylization, robotic agents also commonly follow nonhuman morphologies. Examples of nonhuman appearances include the doglike robot Aibo designed by Sony in 1999 \cite{pransky2001aibo}, a robotic seal designed for therapy in assisted living settings \cite{wada2005psychological}, and Keepon, a robot whose appearance resembled that of a chick \cite{kozima2009keepon}. Finally, robots have also been envisioned as cartoonish characters that blend features from different sources, such as the design of the WALL·E robot by Pixar, a trash compactor with features that suggested humanlike eyes and arms \cite{whitley2012idea}.


In the 1960s, the field of computer graphics and animation started to gain momentum, and by the 1970s most of the building blocks of 3D computer animation were laid, such as surface shading by~\citet{Gouraud:1971} and \citet{Phong:1975} and texture mapping by~\citet{Cat74}. It was not long until computer generated characters began to appear in feature-films such as \textit{Futureworld} (1979, Richard T. Heffron), which was first to showcase a computer animated hand and face, with both wireframe and 3D shading, while the well-known film \textit{Tron} (1982, Steven Lisberger) followed soon after with a whole 15 minutes of computer generated content. Fully animated characters also started to appear in other areas such as music videos (e.g., Mick Jagger’s Hard Woman). 

Ten years later, the technology was developed even further and adopted in films such as \textit{Terminator 2: Judgment Day} (1991, James Cameron) \textit{The Lawnmower Man} (1992, Brett Leonard), and \textit{Jurassic Park} (1993, Steven Spielberg). This was the start of 3D animation receiving a widespread commercial success and it was not long until Pixar Animation Studios released the first entirely computer-animated feature-length film \textit{Toy Story} (1995, John Lasseter). Toy Story was a massive success, largely due to the use of appealing cartoon-characters with plastic appearance, which computer graphics shading was perfectly suited to at that time. 

In the 2000s, more technology was being developed to support the growing industry and Pixar's \textit{Monsters Inc.} (2001, Pete Docter) showed impressive results with simulated fur depicting the subtle secondary motion on the coats of the monster characters. \textit{The Lord of the Rings: The Fellowship of the Ring} (2001, Peter Jackson) pushed new boundaries with realistic crowd simulation, while in the same year \textit{Final Fantasy: The Spirits Within} (2001,  Hironobu Sakaguchi) attempted to create the first photo-realistic virtual humans. While the near-lifelike appearance of the characters in the film was well received, some commentators felt the character renderings appeared unintentionally creepy. Films \textit{The Polar Express} (2004, Robert Zemeckis) and \textit{Beowulf}(2007, Robert Zemeckis) marked further milestones in photorealism, but again received poor audience reactions. Photorealistic rendering was used more successfully for fantasy creatures such as the character Gollum from \textit{The Lord of the Rings: The Fellowship of the Ring}, the first full CGI character in a live-action movie. The actor that drove the movements of Gollum (Andy Serkis) even went on to win the first performance-capture Oscar for his acting in later films. Similar success was achieved with the photorealistic fantasy Navi characters in \textit{Avatar} (2009, James Cameron). 

More recent advancements in 3D scanning, Deep learning, and performance capture have allowed actors to play realistic-depictions of their younger selves (\textit{Bladerunner 2049} (2017, Denis Villeneuve), \textit{The Irishman} (2019, Martin Scorsese), \textit{Gemini Man} (2019, Ang Lee)) or even to play virtual roles after they have passed-away (Peter Cushing in \textit{Star Wars: Rogue one} (2017, Gareth Edwards) and Paul Walker \textit{Fast and Furious 7} (2015, James Wan)).



In the 1980 and 90s, there was also a shift towards interactive media such as games, where real-time animation was employed. This posed new challenges for character creation due to the additional requirements of character responsiveness and agency.

Game characters were thus less visually complex than film characters of the time due to the higher computation cost. The first attempts in the 1980s were in the form of simple 2D sprites such as Pac-Man (Namco), Sonic the Hedgehog (Sega), and Mario (Nintendo). With the advent of home console systems and consumer-level graphics processing units, there was a shift from 2D to 3D in games such as \textit{Quake}, \textit{The Legend of Zelda: Ocarina of Time}, \textit{Tomb Raider}, and \textit{Star Wars Jedi Knight: Dark Forces II}. Characters started to appear more sophisticated and used texture mapping techniques for materials and linear blend skinning for animation. 

In the 2000s, many games utilized cut scenes of cinematic sequences which could achieve higher photo-realism and conversation while disabling the interactive element of the game (e.g., \textit{LA Noire}, \textit{Heavy Rain}, etc.). Nowadays, with real-time raytracing available in game engines, there is no longer a need for photorealism to be restricted to cut-scenes, and we are seeing incredibly realistic depictions of humans and environments in real-time (e.g., \textit{Detroit: Become Human} and \textit{Hellblade: Senua’s Sacrifice}). 

Throughout the years, the graphics and game components have developed rapidly, allowing progressively more realistic depictions every year, though characters with advanced facial animation and conversational capabilities are rarely seen. In commercial games, conversing with non-player characters (NPCs) is usually achieved by selecting predefined conversation texts on the screen, to progress the conversation. There is scope in the future for truly conversational NPCs. Additionally, as virtual reality becomes ever more immersive, we could be about to see the next evolution for the media with higher levels of realism, conversational capabilities and social presence with NPCs.

\section{Design}

\subsection{What is appearance?}
When we say ``appearance'' for agents, we refer to the virtual or physical embodiment that users can experience using their visual faculties. Most agents, from simple static visual representations that accompany chatbots to human surrogates, follow a metaphoric design, that is, the design of the agent takes inspiration or reference from a familiar an existing or envisioned biological entity (e.g., a human, a dog, a grasshopper) or hybrid entity (e.g., a ``trash can'' in appearance, but a cartoonish human in behavior). The expression of a metaphor involves two key dimensions: appearance and behavior. Metaphoric designs can follow consistent or inconsistent implementations across these two dimensions. For example, an agent that follows the metaphor of a dog and appears and behaves like a dog involves a consistent implementation, whereas a dog that speaks involves an inconsistent implementation, integrating dog-like appearance with human-like behavior. The power of agents as a family of computer interfaces comes from metaphoric design, which jumpstarts user mental models and expectations of the system using a familiar representation. For example, a computer system that uses speech as the mode of user interaction and follows a humanlike agent metaphor signals to the user that the system is capable of human mechanisms of communication, such as speech. Similarly, a robot designed to follow the metaphor of a maid or a butler is expected to be competent in household work.

\begin{figure}[t]
  \centering
  \includegraphics[width=\columnwidth]{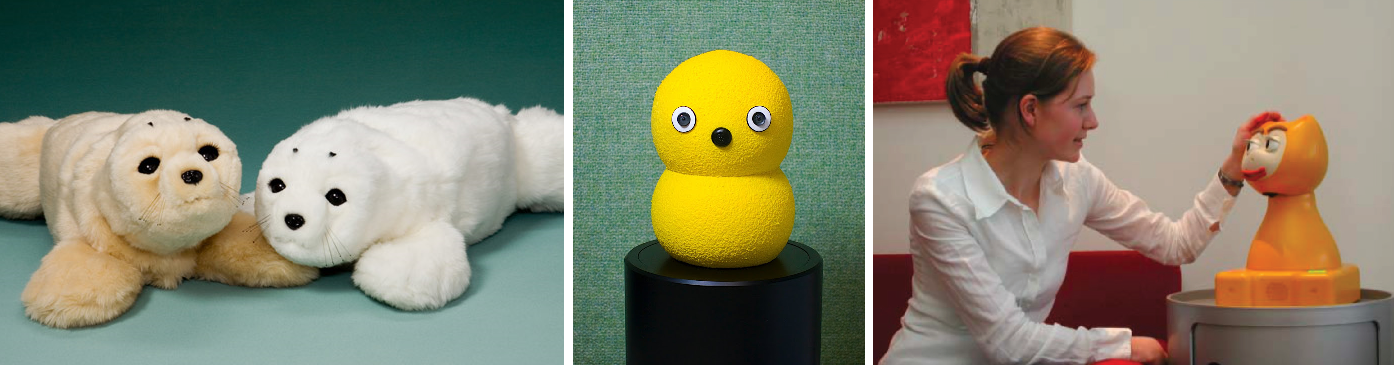}
  \caption{Example metaphors used in the design of robotic agents. \textit{Left to right:} PARO Therapeutic Robot (c) 2014 PARO Robots U.S.; the Keepon robot that followed the metaphor of a chick (c) 2007 BeatBots LLC, \cite{kozima2009keepon}; the iCat robot designed to follow the metaphor of a cat \cite{van2005icat}.}
  \label{fig:metaphors}
\end{figure}

A common approach to designing the appearance of agents is \textit{metaphorical design}, where the design follows a well-known metaphor to elicit familiarity and jumpstart user mental models of the agent’s capabilities. For example, a virtual agent designed to review hospital discharge procedures with patients followed the metaphor of a nurse, appearing on the screen as a nurse in scrubs \cite{bickmore2009taking}. The design of most agents follow a singular metaphor, such as the ASIMO humanoid robot designed to appear as an astronaut wearing a spacesuit \cite{sakagami2002intelligent}, although some designs blend multiple metaphors \cite{deng2019embodiment}, such as the MiRo robot, which integrates multiple animal features chosen to improve perceptions of its friendliness and feelings of companionship \cite{prescott2017miro}. Metaphorical design provides not only morphological features for the design of the agent, but it also provides additional behavioral and physical features such as clothing and environmental context to further support the expression of the metaphor. An example of such features is the design of Valerie the Roboceptionist, a receptionist robot situated in a receptionist’s cubicle, equipped with a backstory that was consistent with the design of the character, and dressed in clothing that was consistent with the backstory and the metaphor that the agent’s design followed \cite{gockley2005designing}. Figure \ref{fig:metaphors} illustrates examples of metaphorical design: the Paro, the Keepon and the iCat robots that followed the metaphors of a seal, a chick and a cat, respectively.

\begin{figure}[t]
  \centering
  \includegraphics[width=\columnwidth]{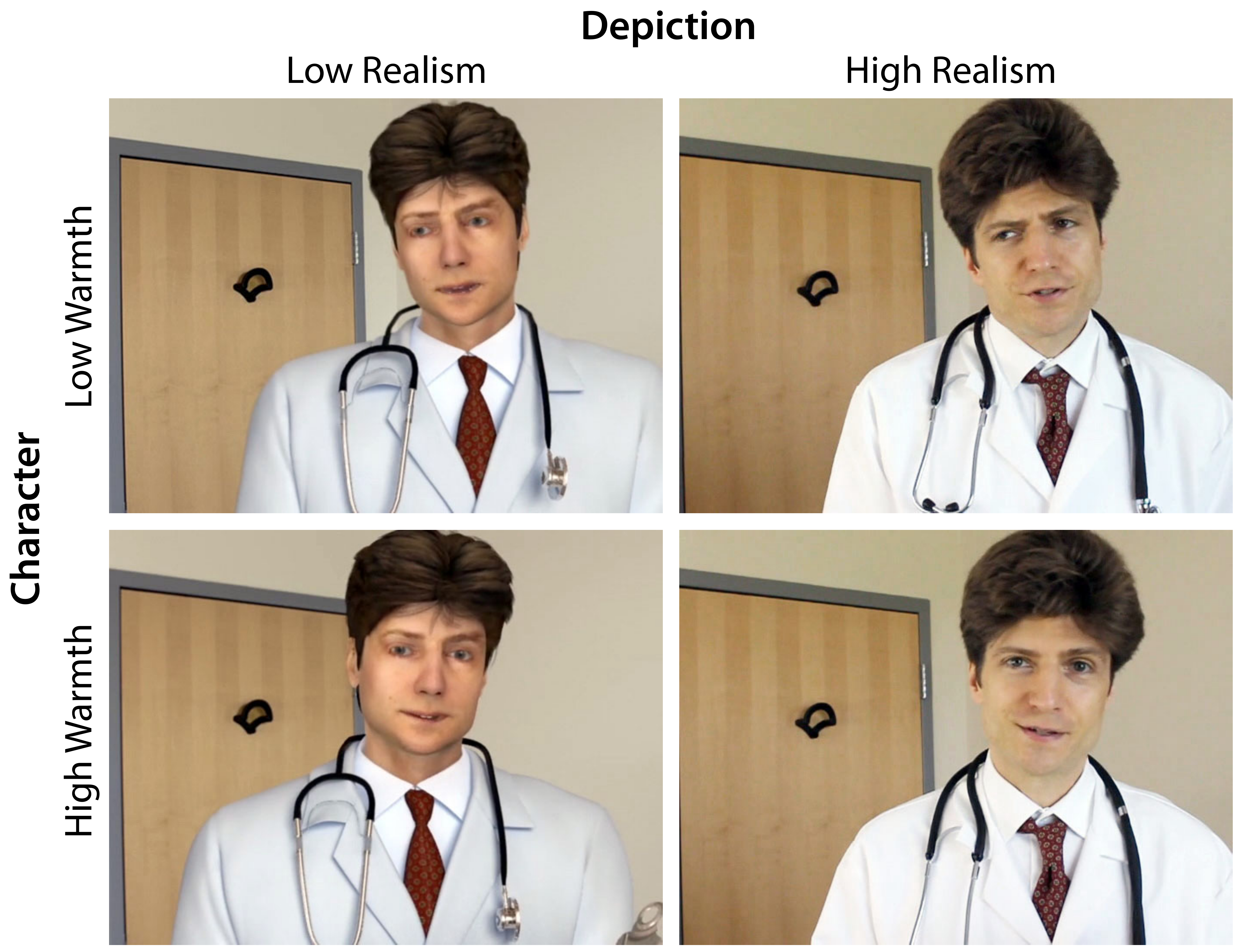}
  \caption{An example of agent design by replicating human experts \cite{Dai:2018}.}
  \label{figure:doctors}
\end{figure}

Virtual agents are also designed to follow different metaphors, most frequently of instructors or experts.  For example, a digital double replica of a real doctor \cite{Dai:2018} was found to be effective at delivering cues of warmth and competence (Figure~\ref{figure:doctors}). More importantly, the virtual doctor’s recommendations also significantly influenced the decisions of participants in the same manner as the real doctor, implying effectiveness at persuasion.

In an educational context, a study on learning outcomes found that a human lecturer is preferable, but that robotic and virtual agents may be viable alternatives if designed properly \cite{Li:2015:VirtualTeacher}. It was also shown that having a stereotypically knowledgeable appearance of the pedagogical agent influenced learning ~\cite{Veletsianos:2010}.

Virtual agents have also been used extensively as assistants. For example, as a navigation assistant in a crash-landing scenario in a study by~\citet{Torre:2018, Torre:2019}, where they had to persuade participants to accept their recommendations about items required for survival. Participants explicitly preferred interacting with a cartoon-like agent than a photorealistic one, and were more inclined to accept the cartoon-agents suggestions. Note that the photo-realistic agent was rated low on attractiveness, and since persuasion and attractiveness have been linked in previous work (e.g., \citet{Pallak:1983}) it may be the case that a more attractive virtual human may have been more persuasive.

Another study compared digital avatars, humans and humanoid robots to determine the influence of appearance on trust and identifying expert advice \cite{Pan:2016}. They found that participants were less likely to choose advice from the avatar, irrespective of whether or not the avatar was an expert. In contrast, experts represented by the robot or by a real person were identified reliably. 



\subsection{Modalities}
\begin{figure}[t]
  \centering
  \includegraphics[width=\columnwidth]{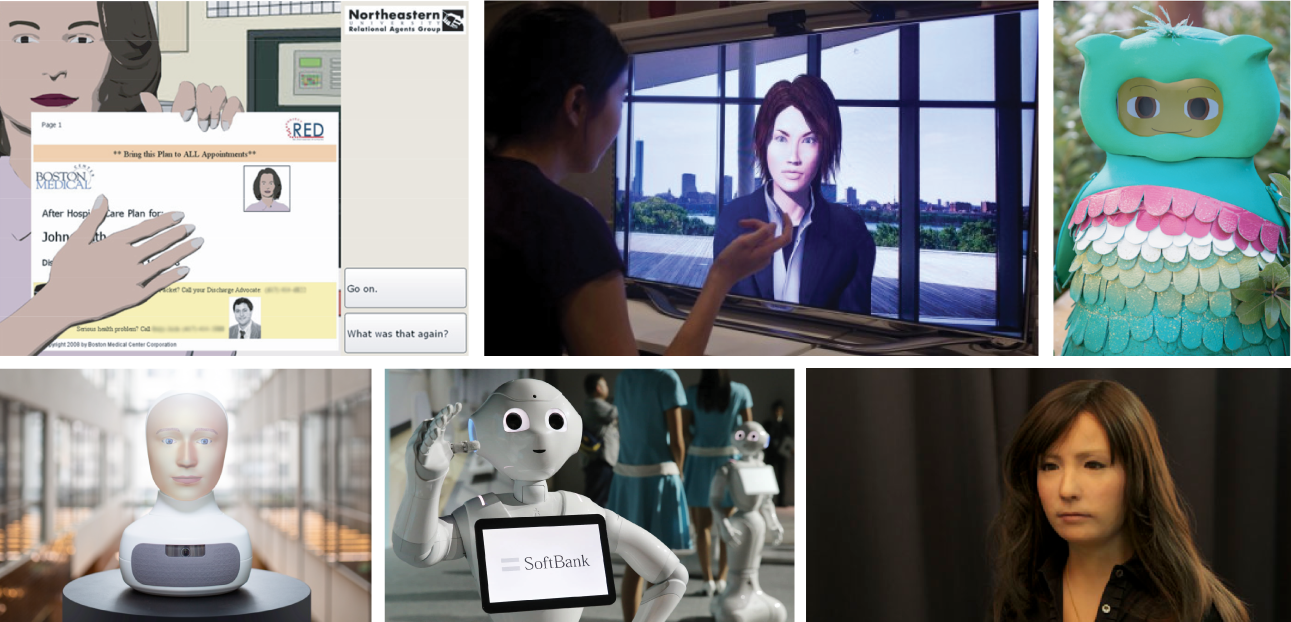}
  \caption{Modalities in which agents are expressed. \textit{Left to right, top to bottom:} the nurse agent Laura rendered as a graphical agent \cite{bickmore2009taking}; the MACH virtual interview coach \cite{hoque2013mach}; the hybrid robot Spritebot with a physical body and a graphical face \cite{deng2019embodiment}; the hybrid FurHat robot with a physical head and a 
projected face (c) 2021 Furhat Robotics; the Pepper physical robot (c) 2021 SoftBank Robotics; the Geminoid F android robot \cite{watanabe2015can}. }
  \label{fig:modalities}
\end{figure}

Appearance can be expressed in graphical, virtual, video-mediated, physical, and hybrid modalities (Figure \ref{fig:modalities}). Agents in graphical modalities are static or dynamic two-dimensional representations, such as a photo, drawing, or animation of a character. For example, ``Laura,'' a virtual nurse designed to support low-literacy patients appeared as a two-dimensional rendering \cite{bickmore2009taking}. Virtual embodiments usually involve three-dimensional simulations that are rendered in real time or replays of rendered animations. An example virtual embodiment is MACH, a virtual interview coach that is rendered in real-time in a virtual environment and presented on a two-dimensional display \cite{hoque2013mach}. Such representations can also be presented in virtual-reality and mixed-reality modalities \cite{garau2005responses}, which provide the user with a more immersive experience of the agent’s embodiment. Agents with a physical appearance involve a robotic embodiment, such as the Robovie robot designed as a shopping mall assistant \cite{iwamura2011elderly} or the Geminoid designed to serve as a human surrogate \cite{nishio2007geminoid}. Users of agents with physical embodiments can also experience the appearance of the agent over video \cite{kiesler2008anthropomorphic}. Finally, hybrid embodiments bring physical and graphical or virtual features together, such as a graphical face appearing on a physical body or graphical features that are projected on the surface of a physical body. Example of hybrid appearances include the FurHat robot \cite{al2012furhat} or Valerie/Tank, a receptionist robot \cite{lee2010receptionist}.

The modality in which an agent is presented affects user perceptions of and experience with the agent. A large body of literature has aimed to compare interaction outcomes across different modalities toward testing the ``embodiment hypothesis:'' that physical embodiment has a measurable effect on user performance and perceptions in interactions with an agent. This body of work shows that, in general, users respond more favorably to agents with stronger embodiments and human-scale sizes. In this context, ``strong'' embodiment refers to modalities that elicit a strong sense of presence, such as physical or hybrid modalities, and ``weak'' embodiment describes modalities such as graphical or virtual that may not elicit a sense of presence at such an extent. \citet{deng2019embodiment} systematically analyzed 65 studies that compared virtual and physical agents in measures of perceptions of the agent and task performance. The analysis showed that 78.5\% of these studies involved improvements in at least one of these categories of measures, consistent with the embodiment hypothesis, 15.4\% involved no change, and 6.1\% involved worsening in at least one of the categories of measures. Among the studies included in this analysis, the most comprehensive comparison was performed by \citet{kiesler2008anthropomorphic}, who compared a collocated robot, a lifesize video projection of a remote robot, a lifesize projection of the virtual version of the robot, and the virtual robot on a computer screen. The measured interaction outcomes generally decreased in this order, the participants responding to the robot more favorably than the virtual agent and the collocated robot more than the projected robot. 

The modality in which the agent is presented not only affects user interaction with the agent, but it also presents different sets of affordances. For example, even if the behaviors of a virtual character and a physical robot are controlled by the same algorithm, the behaviors demonstrated by the agents might look very different due to the differences inherent in the modalities. Unlike virtual characters, physical robots are subject to mechanical limitations and bound by the physical properties of the real world, which might affect the speed with which the agent displays a desired behavior (unbounded in virtual characters, bounded by actuator performance in robots), the sounds that the agent makes (e.g., sound artifacts produced by robots executing motion), the detail with which agent features can be fabricated (bound by modeling and rendering limitations in virtual characters and by physical fabrication limitations in robots), and so on. Physical robots and hybrid agents afford touch interactions and offer texture and material hardness as additional cues. The scale in which the agent is presented is another factor that affects affordances and interaction outcomes. Across all modalities, the closer the agent is presented to human scale, the more likely the agent will support human communication mechanisms. For example, a robot that is expected to be hugged by users must have a size that affords hugging. 

\subsection{Agent Construction}
An important factor that shapes agent appearance is how agents are constructed, which due to historical as well as practical reasons varies based on the modality of the agent. For example, physical agents are constructed using processes and practices from \textit{industrial design}, and their designs are affected by factors such as manufacturing limitations, product safety, and material choice. On the other hand, the construction of virtual characters borrows processes and practices from \textit{animated filmmaking} and \textit{game design}, and their designs are affected by factors including character backstory, the environment in which the agent will be presented, and the mechanisms with which the agent interacts with its environment, the user, and the user’s environment. The paragraphs below outline some of these processes and practices.

\subsubsection{Construction of virtual characters}

Virtual characters have fewer constraints in terms of design than robots, and can be programmed to take on a multitude of different appearances, using a variety of modelling, and rendering techniques. For modelling, virtual characters are typically visualised in 3D using a mesh of consecutive planar polygons which approximate the surface of the human’s body. Polygons are very simple building blocks, and so can be used to describe many diﬀerent shapes. They are also very quick to render on graphics hardware. The construction of 3D models is an established industry with many sophisticated packages available for model-building (e.g., 3D Studio Max, Maya, Blender, Houdini, etc.). Creating detailed 3D virtual characters using these packages is a highly skilled and labour-intensive task primarily due to the fact that 3D models are created using a 2D display and a high level of geometric detail is required to create convincing virtual characters. Generating 3D data for virtual characters can also be accomplished by scanning real people using a range of techniques such as photogrammetry, structured light scanning or laser scanning. Photogrammetry is a type of scanning whereby a collection of still photographs from regular DSLR cameras taken from various angles is all that is required to create a 3D model. Software then analyzes the photographs, matching characteristic points of the object on the images. This creates a point cloud of vertices which can later be converted into a mesh. It is the most commonly used tool nowadays for scanning humans in the visual effects industry, where the number and quality of cameras used in the rig contribute to the accuracy of the recovered mesh. 

3D scanning can also be performed using sophisticated 3D scanning devices to project structured patterns of light or lasers onto the surface of the human to reproduce a 3D model that is a copy of the original. 

For more stylized characters, artists can sculpt characters out of clay and then use one of the mentioned forms of 3D scanning to gather the data onto the computer. 

Professional grade 3D scanners are expensive, but there are also more affordable, consumer-grade technologies such as depth-sensor based 3D scanning (e.g., Microsoft Kinect) and low-cost photogrammetry, which use regular cameras, but results are generally of lower quality and suitable only for low fidelity non-player characters. In the industry, there are a number of rapid character creation products that only require a single photo and create a virtual human within seconds on a tablet or phone \cite{web:didimo, web:itseez3d, web:pinscreen, web:loomai}. These methods are improving in quality and speed with recent advancements in computer vision and deep learning \cite{Hu:2017, thies2016face2face, yamaguchi2018high, saito2017photorealistic, nagano2018pagan}.

\begin{figure}[t]
  \centering
  \includegraphics[width=\columnwidth]{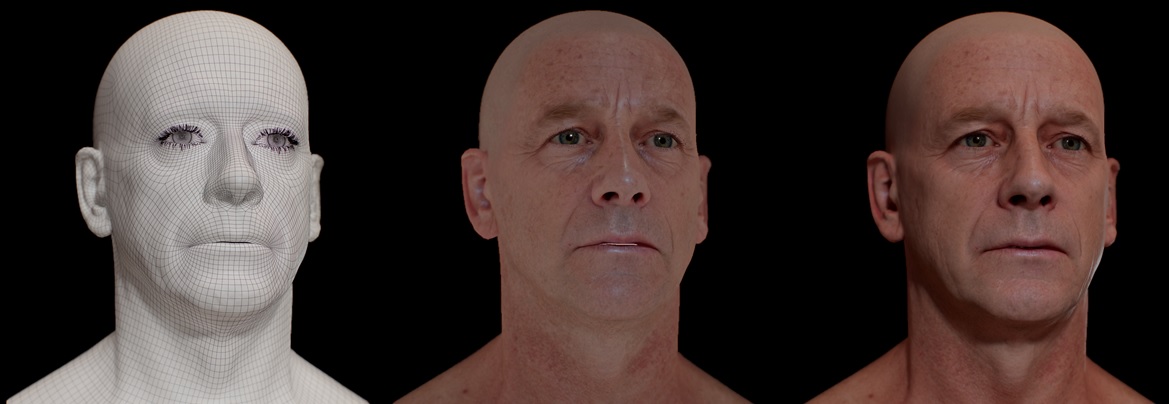}
  \caption{\textit{Left:} Wireframe render of a character with no texture mapping, \textit{Center:} diffuse textures applied, \textit{Right:} high quality rendering including normal maps, specular map, subsurface scattering, global illumination, etc.}
  \label{figure-Wireframe}
\end{figure}


Once a 3D representation of a human character has been created, a number of different techniques can be utilised in order to add detail and realism. A wide variety of render styles from photorealistic to non-photorealistic can be achieved using rasterization for local illumination or ray-tracing for more realistic global illumination \cite{Marschner2016Fundamentals}. While the rasterizer is the current standard for real-time, recent GPU optimization allows for ray-tracing in real-time, and we expect to see much higher realism in virtual characters in the future with global-illumination.

Besides the underlying rendering approach, there are many other methods for adding realism such as texture mapping, and approximating the surface reflectance through shading~\cite{masson2007cg}. Diffuse texture mapping enhances the character by adding image-based information to its geometry, while entailing only a small increase in computation. The basic idea is to map the colour of the image or ‘texture’ onto the corresponding colour of an object at each pixel \cite{Cat74} which adds the illusion of detail to the model, such as clothing material and skin colour.

In order to add colour detail to virtual characters, diffuse texture maps are used which define the color of diffused light (Figure~\ref{figure-Wireframe}). Additionally, there are situations where surfaces are not smooth and roughness needs to be added if it is not present in the geometry. For example, skin is not a smooth surface as it has imperfections such as pores and wrinkles. These details are best added using normal maps which perturb the surface normals to add detail or displacement maps which add geometric detail.

In modern computer graphics, surface properties are governed by shaders, the code snippets describing how a surface should react to incident light. Many physically-based shaders have been developed to produce realistic materials with different Bidirectional Reflectance Distribution Functions (BRDF)~\cite{nicodemus1992geometrical} (the function that relates the incident to the reflected light). More recently, with the rapid advancements in graphics hardware, more complex shading effects approximating a wide range of BRDFs can now be achieved in real-time. For example, subsurface light transport in translucent materials~\cite{Jensen:2001} for realistic scattering of light on the skin was once a technique only used in off-line high-end visual effects, but real-time methods \cite{Jimenez:2009, Jimenez:2010} are now used to enhance the realism in real-time.

Hair for interactive virtual characters has traditionally been modelled using card-based rendering, where images of chunks of the hair are mapped onto large flat sheets, to approximate the shape of a much larger number of individual hairs. Later advancements allowed for modelling each individual hair which dramatically improves realism. For rendering of hair, physically-based fiber reflectance models are used, based on a combination of an anisotropic specular and a diffuse component \cite{Kajiya:1989}. More recently, the scattering distribution of the hair fiber is split into different lobes based on the number of internal reflections within the fiber \cite{Marschner:2003}. 

The use of physically-based simulations is ubiquitous for realism in virtual clothing, where fast mass-spring models~\cite{Liu:2013:FSM} or more complex implicitly integrated continuum techniques \cite{Baraff98} are used in the state-of-the-art. Implementing realistic cloth and hair dynamics in real-time applications still represents a significant challenge for developers since simulation dynamics need to be solved at run-time, and are required to be fast and stable. Based on this, depictions of stiff clothing and hair with little secondary-motion effects are still commonplace for interactive virtual characters across a range of applications from video games to virtual assistants. 


\subsubsection{Industrial design of robots}
\begin{figure}[t]
  \centering
  \includegraphics[width=\columnwidth]{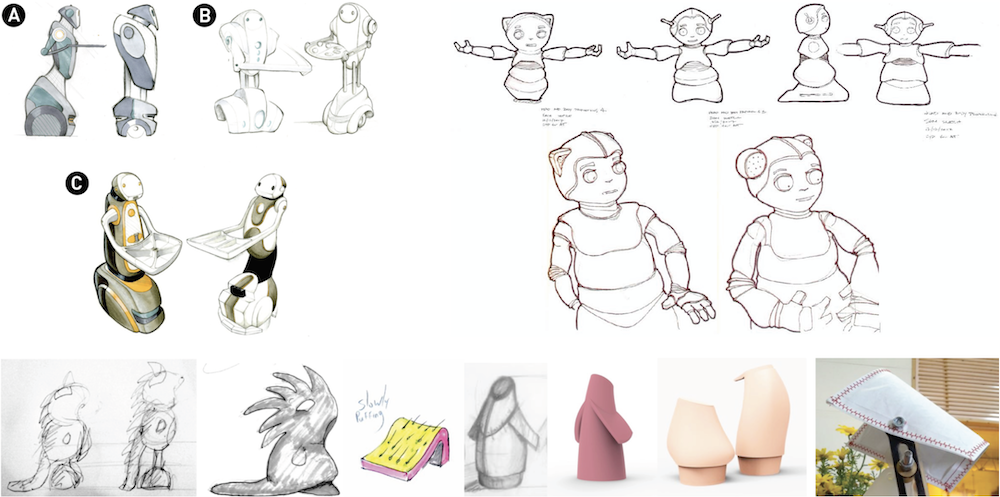}
  \caption{Sketches and models generated during the industrial design of the Snackbot \cite{lee2009snackbot} (top-left), Simon \cite{diana2011shape} (top-right), and Kip1 \cite{hoffman2015design} (bottom) robots.}
  \label{fig:sketches}
\end{figure}

The paragraphs above have discussed design approaches, e.g., metaphorical design, to and the resources used, e.g., facial features, for the development of agent appearance. Another factor that significantly affects agent appearance is the industrial design of physical agents or the physical platforms in which virtual or hybrid agents are presented, including form, material use, scale, color choice, and so on. Although there are no systematic studies of how these factors affect agent appearance or how they must be designed to maximize user experience, the HRI literature includes reports of the design process for the appearance of specific robot platforms. For example, \citet{lee2009snackbot} described the design process for Snackbot, a robot designed to deliver snacks in an academic building, including the form of the housing of the robotic hardware and the snack tray that the robot would carry; the material and colors used to construct the housing and the tray; the height of the robot; and the expressive features of the head and face of the robot. Another example is the design of the Simon humanoid robot, where the research team explored the proportions that the robot's head and body should follow, the placement of the eyes on the head, facial features that would achieve the appearance of a ``friendly doll,'' and the interplay between the design of the housing and structural or mechanical elements of the robot's head \cite{diana2011shape}. \citet{hegel2010social} documented and reported on the industrial design of the social robot Flobi, which included an exploration of the design of the robot's head to follow a ``baby face'' schema; effective color combinations of the robot's face, hair, lips, and eyebrows; and how blushing on the robot's cheeks could be achieved using LEDs placed behind the surface of the face. A final example is the design of Kip1, a peripheral robotic conversation companion, involving form and material exploration through sketches and mock-ups \cite{hoffman2015design}. Figure \ref{fig:sketches} illustrates the sketches and mock-ups generated in the industrial design of some of these examples.

In all of the examples discussed above, the research team engaged professional industrial designers or members of the research team with training in industrial design as well as an iterative design process. The literature does not include any discussion of such considerations for virtual characters, and characters designed for research and commercial use all utilize existing display platforms, such as mobile phones, tablet computers, computer monitors, large displays, or virtual- or mixed-reality environments. Overall, there is a great need for systematic research on the industrial design of the appearance of agents, including the effects of the physical design of the agent itself and the environment within which virtual agents are presented on user interaction and experience.

\section{Features}
The design approaches described above draw on a rich space of features, shaped by the metaphor followed by the design (e.g., humanlike features included in the design of a virtual human), functional requirements of the agent (e.g., light displays placed on physical robots to convey the agent’s status), and/or aesthetic and experiential goals of the design (e.g., material, color, and texture choices for a robot). The paragraphs below provide an overview of this space, focusing on facial and bodily features as well as features that communicate demographic characteristics of virtual and physical agent embodiments.

\subsection{Facial features}
The face of an agent serves as the primary interface between the agent and its user, and facial features make up a substantial portion of the design space for agents. Even when designs lack anthropomorphic or zoomorphic faces, people attribute facial features to them, highlighting the importance of faces in the perception of non-living objects \cite{kuhn2014car}. Designers of virtual and physical agents draw on this human propensity and create faces that can display conversational cues, express affect, and communicate direction of attention.

In order to convey a true feeling of life in a character, the appearance of the eye is highly important. Rendering techniques such as adding specular and reﬂection maps can be very useful for this purpose to increase the appearance of wetness and to reflect the environment. Additionally, more advanced techniques such as ambient occlusion allow for soft shadowing, and refraction to replicate the refraction of light that passes through the eyeball, which is filled with fluid. Creating the geometry of the eye is a difficult task, due to the complexity of the surface but there exist special photogrammetry rigs for capturing the visible parts of the eye---the white sclera, the transparent cornea, and the non-rigidly deforming colored iris 
\cite{Berard:2014}. Computer generated eyes used in computer graphics applications are typically gross approximations of the actual geometry and material of a real eye. This is also true for facial expressions, which typically take a simple approach of linearly blending pre-generated expression meshes (blendshapes) to create new expressions and motion \cite{Anjyo2018}. However, little is known about how these approximations affect user perception of the appearance of virtual characters.

\begin{figure}[!h]
  \centering
  \includegraphics[width=\columnwidth]{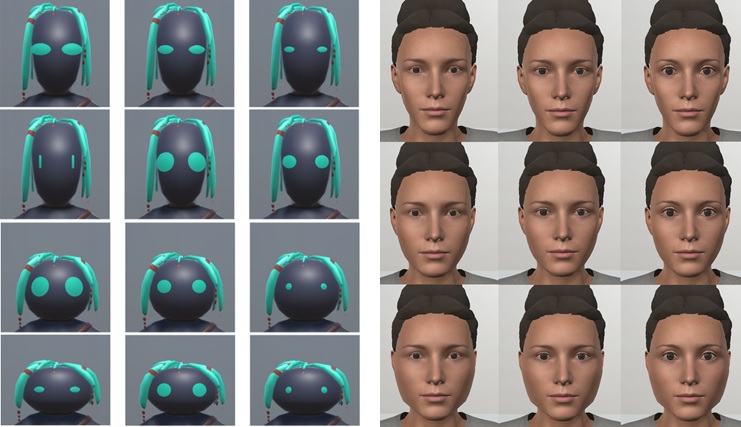}
  \caption{\textit{Left:} Examples of eye and head shape manipulations on abstract characters (based on \cite{Ferstl:2017}), \textit{Right:} More subtle facial feature manipulations on realistic virtual characters (adapted from \cite{Ferstl:2018}). }
  \label{figure-FacialFeatures}
\end{figure}

Similar to the studies on real humans, virtual humans with narrow eyes have been rated as more aggressive and less trustworthy for both abstract creatures \cite{Ferstl:2017} and more realistic depictions \cite{Ferstl:2018} (Figure~\ref{figure-FacialFeatures}). It should be noted that for realistic eye size alterations, the size of the eyes themselves should not be scaled as this will be quickly perceived as eerie and artificial \cite{Wang:2013P}. Instead, the shape of the eyelids can be changed as protruding eyes appear larger, whereas hooded and monolid eyes appear smaller.

In contrast to human face studies, wider faces were not judged as less trustworthy, and were perceived as less aggressive compared to narrow faces for realistic \cite{Wang:2013P} and abstract virtual characters \cite{Ferstl:2017}, even when a particularly masculine rather than a babyface appearance was presented \cite{Ferstl:2018}. The results of these studies support the notion that virtual faces are perceived differently from real human faces. A potential explanation could be the tendency of villains in animated movies to be portrayed with narrow, long, sharp facial features (e.g., Captain Hook in \textit{Peter Pan} (Clyde Geronimi, 1953), Scar in \textit{The Lion King} (Roger Allers, 1994), Maleficent in \textit{Sleeping Beauty} (Clyde Geronimi, 1959)). This tendency could influence the perception of computer-generated characters towards automatic association of narrow faces with dangerous characters.

Other work has addressed the perception of rather unusual facial proportions for realistic characters and their influence on perceived appeal. \citet{Seyama:2007} studied eye size by morphing between photographs of real people and dolls, and found that characters were judged as unpleasant if the eyes had strong deviations from their original size. Participants were more sensitive to the alterations for real faces than for artificial faces. Several studies confirmed that altering facial parts lowers perceived appeal, especially for humanlike characters. \citet{Green08} demonstrated that not only proportions, but also the placement of facial parts may negatively affect the perceived appeal. The measured effect was greater for the humanlike and more attractive faces. Additionally, it has been demonstrated that a mismatch of realism between facial parts negatively affects appeal \cite{MacDorman:2009, Burleigh:2013}. 

Prior work in HRI includes a large body of literature on the facial features of robotic agents. A number of studies aimed to characterize the design space for robot faces. \citet{blow2006art} characterized this space as varying across the dimensions of abstraction, from low to high abstraction, and realism, from realistic to iconic, borrowing from literature on the design of cartoon faces \cite{mccloud1993understanding}. \citet{disalvo2002all} carried out an analysis of 48 robots and conducted an exploratory survey that resulted in a number of design recommendations to improve human perceptions of humanlike robots: (1) the head and the eye space should be wide; (2) facial features should dominate the face with minimal space for a forehead and a chin, (3) the design should include eyes with sufficient complexity; (4) the addition of a nose, a mouth, and eyelids improve perceptions of humanlikeness; and (5) the head should include a skin or a casing that core the electromechanical components. A similar analysis was carried out by \citet{kalegina2018characterizing} of 157 rendered robot faces—physical robots that are equipped with a screen-based face and facial features that are virtually rendered on the screen—who coded the faces for 76 different features and conducted a survey to understand how each feature affected user perceptions of the robot (Figure \ref{fig:facial-features}). The study found that faces with no pupils and no mouth were consistently ranked as being unfriendly, machinelike, and unlikable; those with pink or cartoon-style cheeks were perceived as being feminine; and faces with detailed blue eyes were found to be friendly and trustworthy. Survey participants also expressed preferences for robots with specific facial features for specific contexts of use, e.g., selecting robots with no pupils and no mouth for security work and faces with detailed blue eyes for entertainment applications. Consistently, \citet{goetz2003matching} argued that there is not a universally preferred design for the facial features of a robot, but that people prefer appearances that match the robot's task. They varied the robot's appearance across three stylistic dimensions—human vs. machine, youth vs. adult, and male vs. female—and found that user preferences for facial features presented in these styles depended on the robot's task. In a follow-up study, \citet{powers2006advisor} showed that the length of the robot's chin and the fundamental frequency of its voice predicted whether participants expressed interest in following advice from the robot.

\begin{figure}[t]
  \centering
  \includegraphics[width=\columnwidth]{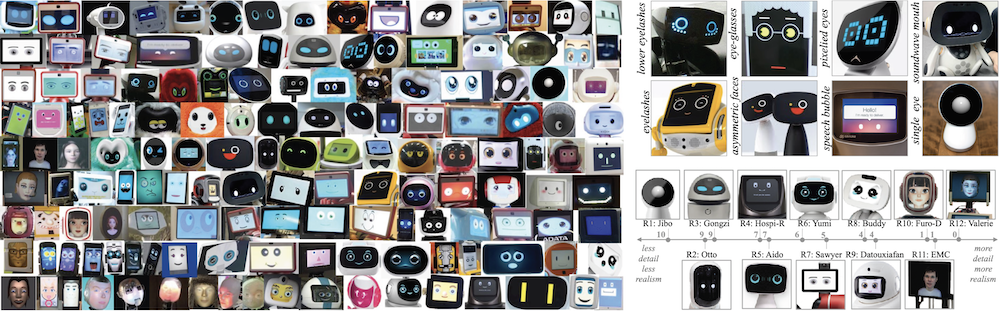}
  \caption{The 157 faces analyzed by \citet{kalegina2018characterizing} (left), their analysis of facial features used in the design of the robot faces (right-top), and the spectrum of facial realism (right-bottom). \textit{Copyright Information:} Images included in this paper under ACM guidelines on Fair Use}
  \label{fig:facial-features}
\end{figure}


The literature also includes reports of the process for the design and development of faces for several robot platforms. For example, the design of the iCub social robot primarily involved the mechanical replication of human anatomical mechanisms to achieve realistic eye and head movements and the design of the rest of the face to follow a ``toy-like'' appearance \cite{beira2006design}. The design specifications for the face of the KASPAR social robot included a sufficiently expressive but minimal design, an iconic overall design (as opposed to a realistic one), a humanlike appearance, and the ability to express autonomy, communicate attention, and display projected expressions \cite{blow2006perception,dautenhahn2009kaspar}. The design of the humanoid robot HUBO integrated an abstract body with the overall appearance of an astronaut and an highly humanlike face using elastomer-based materials that appeared and moved similar to human skin and a 28-degree-of-freedom mechanism to achieve humanlike facial movements \cite{oh2006design}. The faces of robots including the Flobi \cite{lutkebohle2010bielefeld}, Melvin \cite{shayganfar2012design}, and iCat \cite{van2004animation} featured pairs of flexible actuators that served as the robot’s lips and pairs of eyebrows to express emotion. As discussed earlier, the design of the face of the Flobi robot, shown in Figure \ref{fig:facial-features}, additionally included sophisticated mechanisms for emotion expression, such as lights placed behind the cheeks to enable the appearance of blushing. These reports illustrate how different facial features come together in the design of different robot systems and point to specific examples in the design space of facial features for robots.

\subsection{Bodily features}
While the face serves as the primary interface for human-agent interaction, the remainder of an agent’s body also contributes to the appearance of the agent. The design space for an agent’s body primarily includes several bodily features, how these features come together structurally, and how they are represented.

A virtual agent's body can be presented in a range of different styles, from low-detailed stick-figures or point-light displays to photorealistic bodies or anthropomorphised creatures, and there have been some studies aimed at investigating the effect of the body representation on perception of the agent’s appearance and actions. Most studies apply motion captured animations to a virtual character and map the motion onto a range of bodies and assess if the different bodies change the meaning of the motion. Typically, factors such as emotion, gender, and biological motion are chosen since these have all been shown to be identifiable solely through motion cues (e.g.,\cite{Johansson73,Cutting77,Kozlowski77}) thus allowing the contribution of the bodies appearance to be assessed.

Beginning with a study by \citet{Hodgins:1998}, the amount of detail in a virtual character's representation has been studied to investigate the effect on perception. Their study found that viewers’ perception of motion characteristics is affected by the geometric model used for rendering. They observed higher sensitivity to changes in motion when applied to a polygonal model, than a stick figure. \citet{Chaminade:2007} also found an effect on motion perception, where character anthropomorphism decreased the tendency to report their motion as biological, while another study found that emotions were perceived as less intense on characters with lower geometric detail \cite{Mcdonnell09b}.

\begin{figure}[t]
  \centering
  \includegraphics[width=\columnwidth]{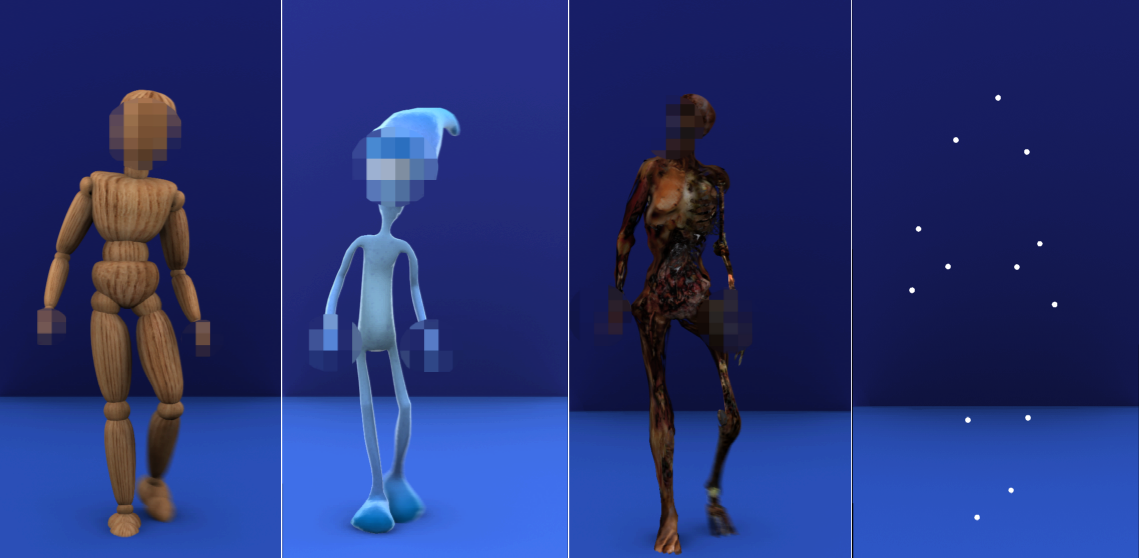}
  \caption{Different structural and material representations for agent body \cite{McDonnell:2009}.}
  \label{figure-Zombie}
\end{figure}

\begin{figure}[!bh]
  \centering
  \includegraphics[width=\columnwidth]{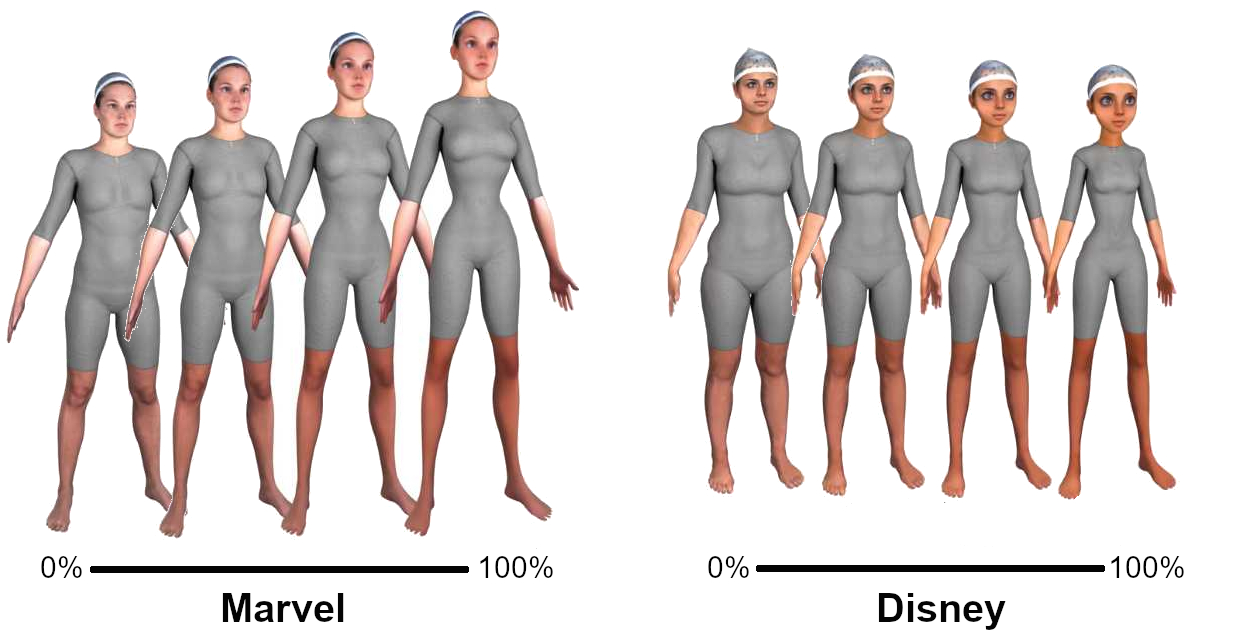}
  \caption{Stylization applied at different levels (33\%, 66\%, 100\%) to captured performer body (0\%) in Marvel and Disney styles (image based on \citet{Fleming:2016}).}
  \label{figure-Fleming}
\end{figure}

Body shape has also been investigated where it was found that a virtual character’s body does not affect recognition of body emotions, even for extreme characters, such as a zombie with decomposing flesh \cite{Mcdonnell09b} (Figure~\ref{figure-Zombie}). \citet{Fleming:2016} evaluated the appeal and realism of female body shapes, which were created as morphs between a realistic character and stylized versions following design principles of major computer animation studios. Surprisingly, the most appealing characters were in-between morphs, where 33\% morphs had the highest scores for realism and appeal and 66\% morphs were rated as equally appealing, but less realistic (Figure~\ref{figure-Fleming}).

The perception of sex of a virtual character’s walking motion has also been found to be affected by body shape. Adding stereotypical indicators of sex to the body shapes of male and female characters influences sex perception. Exaggerated female body shapes influenced sex judgements more than exaggerated male shapes~\cite{McDonnell09a}.

In virtual reality, embodiment of virtual characters is where the user is positioned virtually inside the body of a virtual avatar, where they have agency over that virtual body. The character model used for the virtual avatar can affect the behaviour of the user, from becoming more confident when embodied in a taller avatar, more friendly as an attractive avatar~\cite{Yee2009}, to reducing implicit racial bias by embodying an avatar of a different race \cite{Banakou:2016}. This powerful effect is referred to as the Proteus effect \cite{Yee:2007Proteus} (named after the Greek god known for his ability to take on many different physical forms). The use of self-avatars or virtual doppelgangers has also been shown to affect outcomes, with generally a positive influence on aspects such as cognitive load \citep{Steed2016}, pain modulation \citep{Romano2014} and embodiment \citep{Kilteni:2012, Fribourg:2020}. These effects describe to some extent the dynamism of interactions between users and avatars.

\begin{figure}[t]
  \centering
  \includegraphics[width=\columnwidth]{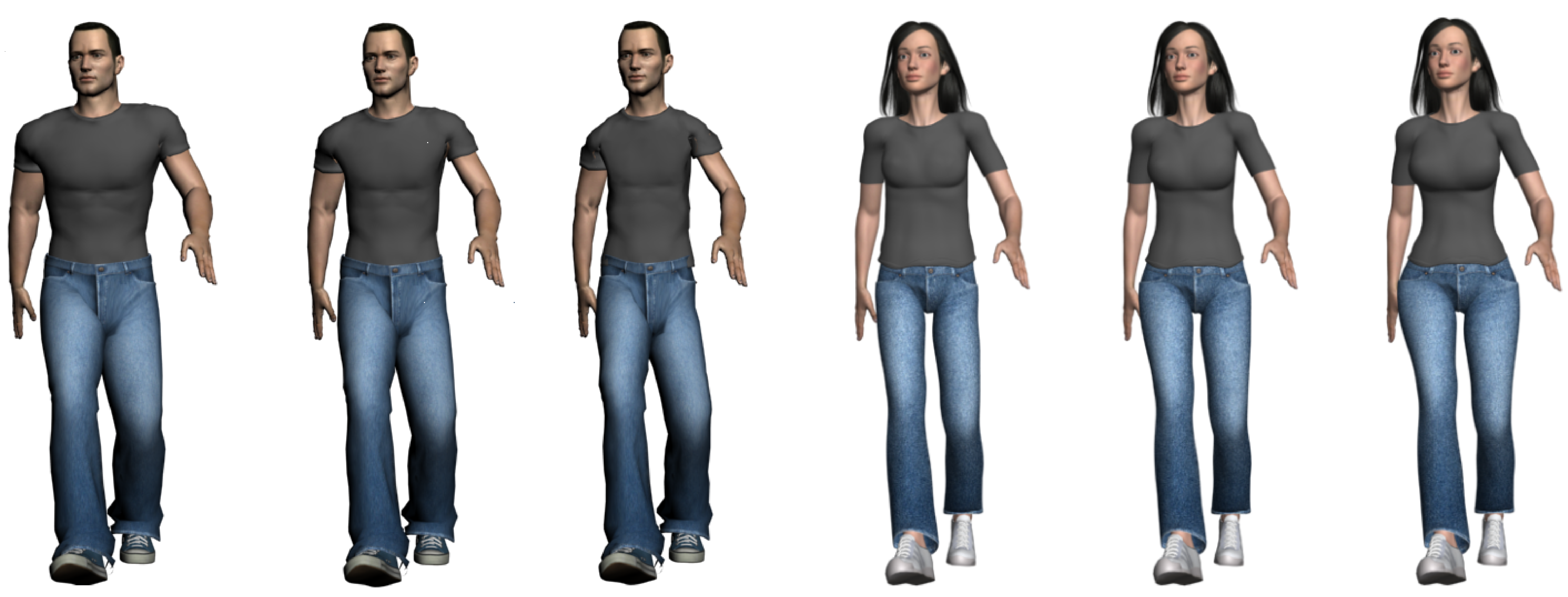}
  \caption{Six body shapes with indicators of gender \cite{McDonnell09a}.}
  \label{figure-Gender}
\end{figure}


The design of a physical robot’s body is shaped by a number of factors, including the metaphor that the design follows, the functional requirements of the robot, and environmental constraints that the design must consider. The first factor, the design metaphor, might dictate how the body of the robot is structured and the features that are articulated in the design. For example, the Paro robot \cite{wada2007living} follows the metaphor of a baby seal, and the design of the robot’s body roughly follows the form of a seal, including fore and hind flippers. The functional requirement of the robot might include specific forms of mobility, such as holonomic movement, climbing stairs, or movement across rough terrain, or prehensile manipulation involving a single arm or two arms. Depending on such design requirements, the design of the body of a robot might follow a humanoid design including humanlike limbs attached to a torso, such as the ASIMO robot \cite{sakagami2002intelligent}, or a single arm attached on a mobile base, such as the Fetch robot \cite{wise2016fetch}. Finally, the environment that the robot is designed for can dictate the bodily features of the robot, such as requiring that a robot that crawls into tight spaces has a low profile and limbs that can be tucked away, such as a Packbot robot \cite{yamauchi2004packbot} used in search-and-rescue scenarios.

In addition to bodily features borrowed from the design metaphor, such as the hind flippers of a seal or the legs of a human, the design of physical robots also utilize features that facilitate specific functions. These functions include communication, and features that support communication include lights that communicate the robot’s affective states using different colors \cite{bethel2007survey} or light arrays that convey information about the robot’s direction of motion using light patterns \cite{szafir2015communicating}. Features of a robot’s body may also support transferring items, such as a tray that the Snackbot robot held to carry food items \cite{lee2009snackbot} and the different configurations of carts that hospital delivery robots pull to transport materials \cite{ozkil2009service}.

\begin{figure}[!b]
  \centering
  \includegraphics[width=\columnwidth]{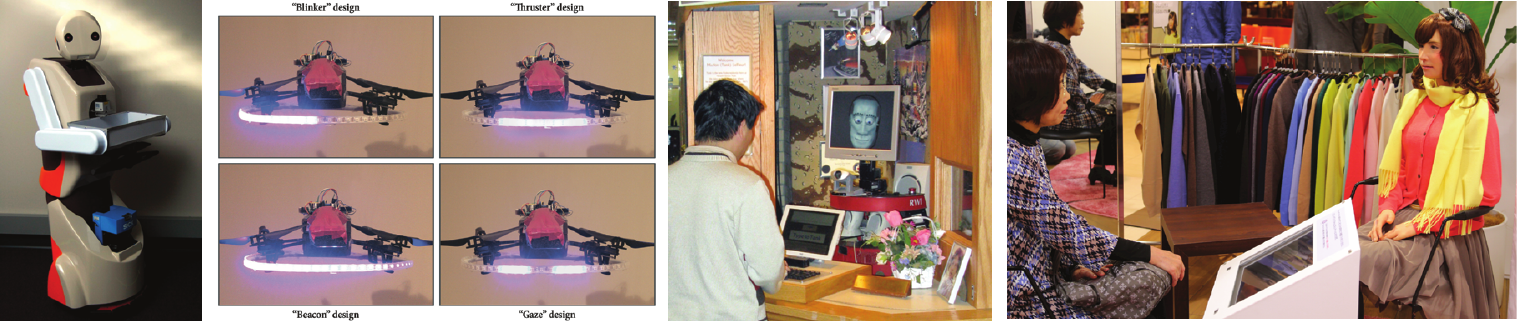}
  \caption{Bodily features that support specific functions, such as a tray that the robot uses to delivery snacks \cite{lee2009snackbot} (left) and light arrays that a flying robot uses to communicate direction \cite{szafir2015communicating} (left-center), and that support the agent's character, such as a booth and clothing for a receptionist robot \cite{lee2010receptionist} (right-center) and clothing for a surrogate robot \cite{watanabe2015can} (right).}
  \label{fig:body-features}
\end{figure}

An agent’s body can also include bodily features, such as clothing or furniture, designed to support the agent’s character or backstory or eventually improve user experience with the agent. For example, the Roboreceptionist robot was placed in a booth that resembled an information booth and wore clothes that were consistent with the gender and the backstory of its character \cite{gockley2005designing}. The Geminoid robot, a highly realistic android developed to serve as a robotic surrogate to support remote communication, was constructed to resemble its creator and dressed in similar fashion \cite{nishio2007geminoid}. Figure \ref{fig:body-features} illustrates examples of bodily features that support specific functions, such as a tray, and that support the agent's character, such as clothing.

\subsection{Features expressing demographic characteristics}
Agent appearance communicates other attributes of the character of the agent, such as gender, age, race, and ethnicity. Virtual agents are usually designed as distinctive characters, such as the two female nurse characters, one middle-aged Caucasian and one middle-aged African American, designed by \citet{bickmore2009taking} to match user patient demographics. Physical agents, on the other hand, are designed as characters with ambiguous features and interchangeable parts that highlight specific character attributes, such as the interchangeable hair and lips of the Flobi robot that communicate a male or female gender \cite{lutkebohle2010bielefeld} (Figure \ref{fig:facial-features}).

A large body of research on human-agent interaction has shown such character attributes to significantly shape interaction outcomes. For example, \citet{siegel2009persuasive} asked participants to make an optional donation to a robot that used pre-recorded male or female voices, which research has shown to be sufficient to trigger gender stereotypes \cite{nass1997machines}, and found a significant interaction between robot and participant gender over the proportion of participants who donated any amount, e.g., men consistently donating more to a female robot. \citet{eyssel2012s} manipulated the gender of the Flobi robot by varying the robot’s appearance via its interchangeable parts for hair and lips and found that participant perceptions of the male and female robots closely followed gender stereotypes. The male robot was perceived as having more agency and being more suitable for stereotypically male tasks (e.g., repair), and the female robot was perceived as being more communal and being more suitable for stereotypically female tasks (e.g., childcare). 

\begin{figure}[t]
  \centering
  \includegraphics[width=\columnwidth]{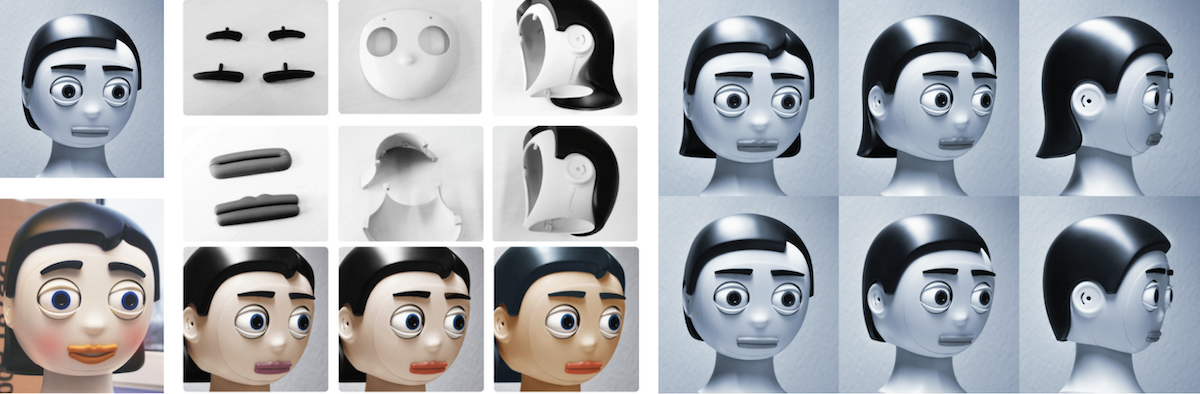}
  \caption{Facial features of the Flobi robot that provide the robot with different demographic characteristics. \textit{Left:} neutral male (top) and smiling female (bottom) faces; \textit{Center:} the physical parts that represent facial features; \textit{Right:} different hair and lip styles. Adapted from \citet{lutkebohle2010bielefeld}.}
  \label{fig:facial-features}
\end{figure}


The effect of stereotypes has also been studied for virtual characters, mostly in the context of embodiment in virtual reality. The Proteus Effect, as mentioned previously, has additionally shown that users conform to stereotypes associated with their avatar’s appearance. For example, embodiment in female avatars made players more likely to conform to female-typed language norms \cite{Palomares:2010} and made them more likely to engage in healing activities \cite{Yee:2011}. Interestingly, these effects were observed regardless of the actual gender of the player, indicating a tendency to conform to expectations associated with the virtual gender.

In other work, \citet{Zibrek15} explored gender bias on different types of emotions applied on male and female virtual characters. They found that emotion biases gender perception according to gender stereotypes: an angry motion is seen as more male, while fear and sadness are seen as less male motions, and they observed a contrast effect where anger was seen as more male when viewed on a female model than when viewed on a male model. Similar effects were found for real humans \cite{Hess04}, indicating that virtual humans follow similar stereotyping effects.

\subsection{Realism, Appeal, Uncanny Valley}
Metaphorical design involves the application of a familiar metaphor to the design of an agent, such a virtual human following the metaphor of a human. In practice, metaphors are applied at different levels of abstraction due technical limitations (e.g., inability to closely replicate the original metaphor) and design choices (e.g., stylization). \citet{deng2019embodiment} argued that designs follow discrete metaphors (e.g., a ``baby seal'' metaphor) but the realism in which these metaphors are applied to vary along a spectrum of abstraction (e.g., a stylized or abstract household robot vs. a highly realistic robotic surrogate). The design choices of metaphor and abstraction result in differences in user perceptions of the agent and experience with it.

In the classic textbook ``Disney Animation: The Illusion of Life,'' ~\citet{Thomas1995illusion} use the term \emph{appeal} to describe well designed, interesting and engaging characters. This is contrary to many face perception studies, which use the term appeal and attractiveness interchangeably. Appeal is an essential ingredient for virtual characters in video games and movies, as well as for avatars, agents, and robots, to ensure audience engagement and positive interactions. Creating highly detailed, photorealistic virtual characters does not necessarily produce appealing results \cite{Geller:2008}, and it is often the case that more stylized approximations evoke more positive audience responses and engagement \cite{Zell:2019}. However, additional factors are the context of the interaction and how appropriate the appearance is under the circumstances. For example, having a fun cartoon-appearance may be less appropriate for a more serious application such as a for a business meeting \cite{Junuzovic:2012} or medical training \cite{Volante16}, etc. Perception of appeal of virtual characters is an ongoing area of research, with the ultimate goal to speed-up or automate the process of producing appealing characters, and avoid negative reactions from audiences.

The term Uncanny Valley (UV) is often used to describe the negative reactions that can occur towards virtual characters. It is a feeling of repulsion produced by artificial agents that appear close to human-form but not quite real. This UV phenomenon was first hypothesized by in the 1970s by robotics professor \citet{Mori:1970}. Mori's original hypothesis states that as a robot’s appearance becomes more human, humans evoke more positive and empathetic responses, until a point where the response quickly becomes strongly negative resulting in feelings of disgust, eeriness and even fear. Once the robot’s appearance becomes less distinguishable from a human being, the emotional response becomes positive once again. This negative response has been attributed to many causes such as motion errors or lack of familiarity or a mismatch in realism between elements of character design. More recently, the UV hypothesis has been transferred to virtual humans in computer graphics, and has been explored directly in some studies \cite{MacDorman:2009, Bartneck:2009}. Virtual faces in particular are difficult to reproduce as humans are very adept at perceiving, and recognising other faces and facial emotions. 

\begin{figure}[t]
  \centering
  \includegraphics[width=\columnwidth]{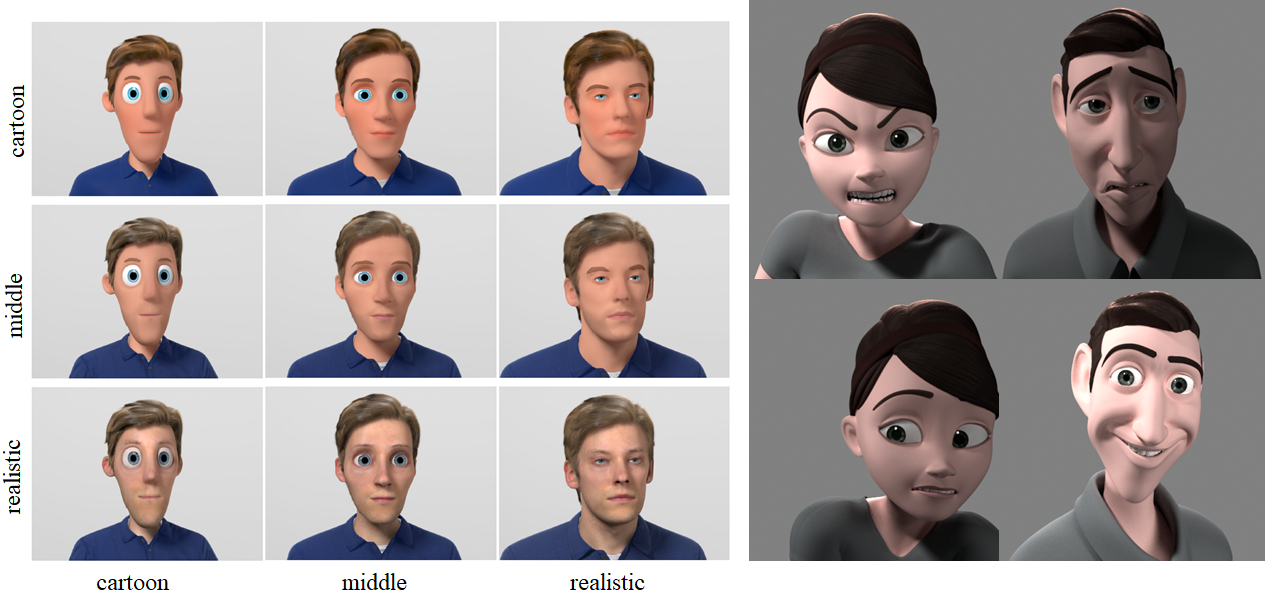}
  \caption{\textit{Left:} Examples of manipulating material (y-axis) and shape (x-axis) to vary character realism and appeal, image based on \citet{Zell:2015}, \textit{Right:} Examples of brightness and shadow alterations on cartoon characters displaying emotion which were shown to change the perceived intensity of emotion \cite{Wisessing:2020}.}
  \label{figure-ZellWisessing}
\end{figure}


As discussed previously, the appearance of a character can be separated into texture, materials, shape and lighting. Various studies have attempted to isolate these factors and independently examine the effect on appeal and UV. 

\citet{Wallraven:2007:ERC} studied the perceived realism, recognition, sincerity, and aesthetics of real and computer-generated facial expressions using 2D filters to provide brush, cartoon, and illustration styles and found that stylization caused differences in recognition accuracy and perceived sincerity of expressions. Additionally, their realistic computer-generated faces scored high aesthetic rankings, which is contrary to the UV theory. \citet{Pejsa:2013} additionally found no effect on appeal or lifelikeness between a character with human proportions and one with stylized geometry including large eyes, while other studies found realistic and cartoon depictions to be equally appealing when expressing personality \cite{Ruhland15} and when a user had agency over their movements~\cite{Kokkinara:2015}.

In order to investigate the effect of stylization in more detail, \citet{McDonnell:2012} created a range of appearances from abstract to realistic by altering the rendering style (texture, material and lighting) of a realistically modelled male character while keeping the shape and motion constant. They analyzed subjective ratings of appeal and trustworthiness and found that the most realistic character was often rated as equally appealing or pleasant as the cartoon characters, and equally trustworthy in a truth-telling task. A drop in appeal occurred for characters in the middle of the scale (rated neither abstract nor realistic), which was attributed to the difficulty in categorizing these characters due to their uncommon appearance \cite{Saygin2012}. Other studies of the UV that used still images generated by morphing between photographs and animated characters also found valleys in participant ratings of uncanniness for intermediate morphs \cite{Hanson:2005, Seyama:2007, Green08}. This idea was further developed in the categorization ambiguity hypothesis~\cite{Yamada:2013, Cheetham13}, where it was shown that this response is more prominent when the morph is between a real human and an inanimate object or representation of a human. Studies focusing on neurocognitive mechanisms attribute negative evaluation to a competing visual-category representations during recognition \cite{Ferrey:2015}. 

This effect was also investigated in a study by \citet{Carter13} where they created a realistic, cartoon, and robot female character and assessed subjective pleasantness ratings as well as analyzing eye-tracking as a psychophysiological measure. Contrary to the UV theory, they found higher ratings of unpleasantness for their cartoon than for their realistic character, and that fixations were affected by subjective perceptions of pleasantness.

Investigating yet more parameters, \citet{Zell:2015} independently examined the dimensions of shape, texture, material and lighting, by creating a range of stimuli of characters with various levels of realism and stylization (Figure \ref{figure-ZellWisessing} (left)). Their study identified that the shape of the character’s face is the main descriptor for realism, and material increases realism only for realistic shapes. Also, that strong mismatches in stylization between material and shape made characters unappealing and eerie, in particular abstract shapes with realistic materials were perceived as highly eerie, validating the design choices of some horror movies with living puppets. Finally, blurring or stylizing a realistic texture can achieve a make-up effect, increasing character appeal and attractiveness, without reducing realism. The opposite was found in a study on body stylization, where the stylization of body shape predicted appeal ratings rather than improvements to render quality \cite{Fleming:2016}.

\begin{figure}[t]
  \centering
  \includegraphics[width=\columnwidth]{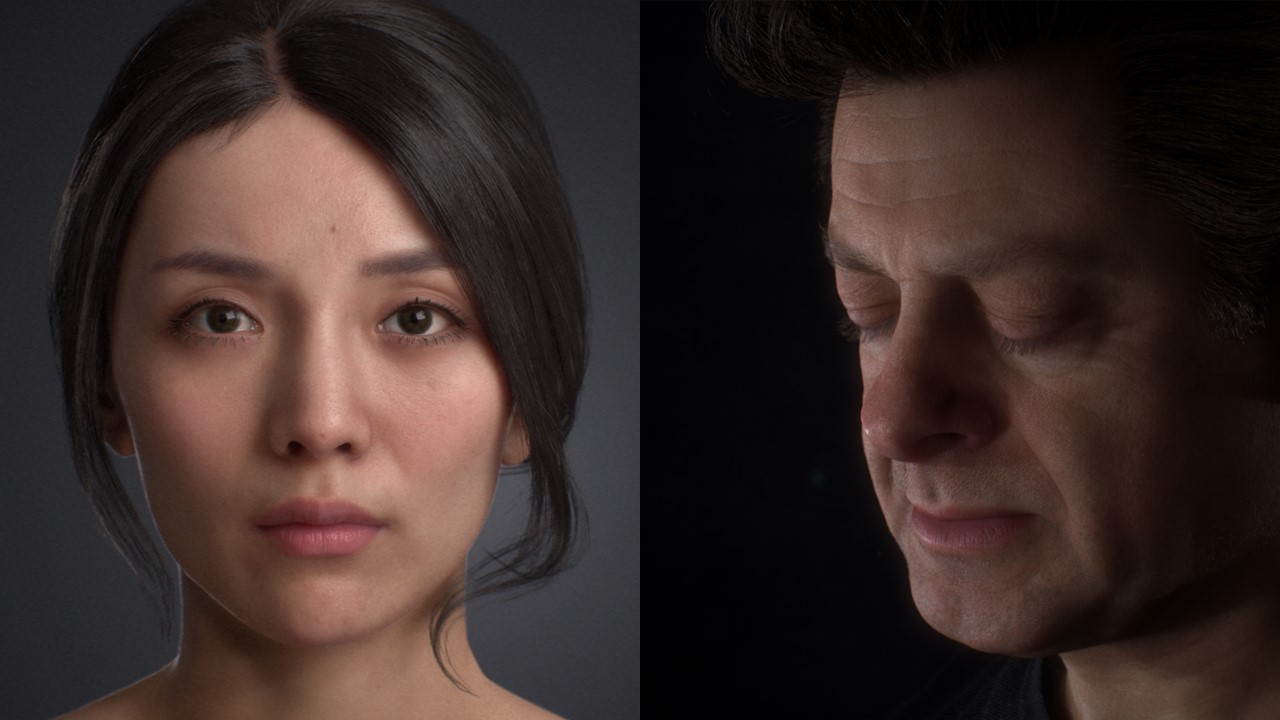}
  \caption{State-of-the-art real-time virtual humans in Unreal Engine 4 created by 3Lateral in collaboration with Cubic Motion, Epic Games, Tencent and Vicon. \textit{Left:} Siren demo. \textit{Right:} virtual replica of the actor Andy Serkis. With permission of Epic Games \copyright~2020.}
  \label{figure-Epic}
\end{figure}

More recently, \citet{Wisessing:2020} carried out an in-depth analysis of the effect of lighting on appeal, particularly brightness and shadows, and found that increasing the brightness of the key-light or lessening the key-to-fill ratio (lighter shadows) increased the appeal ratings (Figure \ref{figure-ZellWisessing} (right)). They also found little effect of key-light brightness on eeriness but reported reduced eeriness as a consequence of lightening the shadows, which could be used to reduce UV effects of virtual characters. However, shadow lightening did not improve appeal for characters with realistic appearance, and thus key-light brightness alone should be used to enhance appeal for such characters.

Several studies in immersive VR have also examined the effect of character appearance on viewer responses, focusing on co-presence, i.e., the sense that one is present and engaged in an interpersonal space with the character~\cite{Biocca97,Garau03}. While some evidence confirms the importance of realistic appearance~\cite{Nowak01,Zibrek:2019}, others put less importance on it~\cite{Slater02,Garau03}. On the other hand, a mismatch between the realism of behaviour and appearance has been often shown to lower the feeling of co-presence~\cite{Bailenson05}. There are a number of reasons why mismatches may cause negative effects on the viewer. A mismatch between the physical and emotional states of a character violate expectations and thus can result in a breakdown in how users experience agents \cite{Vinayagamoorthy:2006}.

\section{Summary}


Technical advancements are increasingly pushing the boundaries of how agents are designed and developed, the capabilities of these agents, and their use in human environments. The rapid development in real-time rendering technologies has enabled incredibly detailed, high-quality virtual character appearances (Figure~\ref{figure-Epic}), often reaching photorealism~\cite{seymour2017meet, 3lateral_siren_2018}. Deep learning is also improving the ease and speed at which characters can be created, even from a single photograph~\cite{yamaguchi2018high}. Additionally, animation and behaviours are starting to become easier and less expensive to create, allowing virtual human technologies to be more accessible to a wider audience than ever before. With these advancements comes the increasing use of characters across different domains such as education, sales, therapy, entertainment, social media, and virtual and augmented reality. 

New methods are also emerging for the construction of physical robots. Rapid fabrication methods, such as 3D printing, have led to the development of new robot morphologies, including 3D printable robots inspired by ``origami'' \cite{onal2014origami} and robots with soft skin that can change appearance and texture to communicate internal states to the user \cite{hu2018soft}. Mixed-reality technologies are also being utilized to facilitate human interaction with robots, including displaying cues that communicate the motion intent \cite{walker2018communicating} and and the field of view \cite{hedayati2018improving} of the robot. Finally, robots are increasingly being integrated into human environments across different domains, including manufacturing \cite{sauppe2015social}, education \cite{belpaeme2018social,michaelis2018reading}, food services \cite{jennings2019study}, hospitality \cite{tussyadiah2018consumer}, surveillance \cite{inbar2019politeness}, and healthcare \cite{mutlu2008robots,miseikis2020lio}. As applications proliferate, we will gain a better understanding of how the design space for agent appearance is utilized to support each application domain, how the features of this space affect user perceptions of and experience with these agents, and how the appearance of robotic agents might be designed to support personalization, customization, and environmental fit.

In this chapter, we have shown that the choice of appearance can have implications for human interactions in a number of ways, including changes to the perception of personality, emotion, trust, and confidence. Studies have shown that the many factors that constitute the final appearance of an agent, such as the design metaphor, modality of representation, and methods of agent construction, including modelling, texturing, materials, and even lighting, have different effects on how people perceive and respond to it. This multidimensionality has the drawback that some factors might cancel each other out or amplify each other, leading to inconsistent conclusions. Additionally, more frequent exposure to agents and increasing technological sophistication may continuously change the way we perceive them, much like how we are becoming more and more sensitive to poor visual effects in movies~\cite{Tinwell:2011}. The need for understanding the implications of different appearances of agents has therefore never been greater.